\documentstyle[12pt]{article}
\begin{document}

\noindent

\vspace{.5cm}
\begin{center}
{\large\bf ANALYTICAL STUDY OF THERMONUCLEAR REACTION
PROBABILITY INTEGRALS}
\\[1cm]
{\bf M.A. Chaudhry}\\
\end{center}
Department of Mathematical Sciences, King Fahd University of Petroleum and Minerals, Dhahran 31261, Saudi Arabia \\[2mm]
\begin{center}
{\bf H.J. Haubold}\\
\end{center}
Outer Space Office, United Nations,Vienna International Centre, P.O. Box 500, 1400 Vienna, Austria\\
\begin{center}
and\\[2mm]
{\bf A.M. Mathai}\\
\end{center}
Department of Mathematics and Statistics, McGill University, 805 Sherbrooke Street West,\\
Montreal, Quebec, Canada H3A 2K6\\

\vspace{.25cm}
\noindent
{\bf Abstract.} An analytic study of the reaction probability integrals
corresponding to the various forms of the slowly varying
cross-section factor  $S(E)$ is attempted. Exact expressions for
reaction probability integrals are expressed in terms of
the extended gamma functions.

\vspace{2cm}

\newpage

\baselineskip=26pt

\begin{center}
{\bf 1. \, Introduction}
\end{center}
\vspace{8mm}

\renewcommand{\thesection}{1}
\renewcommand{\theequation}{\thesection.\arabic{equation}}
Nuclear reactions govern major aspects of the chemical evolution of
galaxies and stars (Fowler, 1984, Morel et al., 1999).  The analytic
study of the reaction rates and
reaction probability integrals was undertaken by Critchfield (1972),
Anderson et al. (1994), Haubold and Mathai (1998), and
Chaudhry (1999).  A proper understanding of the nuclear reactions
that are going on in hot cosmic plasma, and those in the laboratories
as well, requires a sound theory of nuclear-reaction dynamics
(Clayton, 1983, Bergstroem et al., 1999).  The rate $r_{ij}$ of unlike
reacting nuclei $i$ and $j$ in the case of nonrelativistic nuclear
reactions taking place in nondegenerate environment is expressed as
(Clayton, 1983, Lang, 1999)
\begin{equation}
r_{ij} = n_i n_j \left(\frac{8}{\pi \mu}\right)^{1/2}
\left(\frac{1}{kT}\right)^{3/2} \int^\infty_0 E\sigma(E)e^{-E/KT}dE,
\end{equation}
where $n_i$ and $n_j$ denote the particle number densities of the
reacting nuclei $i$ and $j$, $\displaystyle \mu = \frac{m_i m_j}{m_i + m_j}$
is the reduced mass of the reacting nuclei, $T$ is the temperature,
$k$ is the Boltzmann constant, $\sigma(E)$ is the cross-section for the
reaction under consideration, and $v = \left(\frac{2E}{\mu}\right)^{1/2}$ is
the relative velocity (for reactions between like nuclei $i$ = $j$, $n_in_j$ has to be replaced by $n^2$).  Thus (Clayton, 1983, Lang, 1999)
\begin{equation}
r_{ij}=n_in_j \lambda:=n_in_j\langle \sigma v\rangle:= \int^\infty_0 \sigma(E)v(E)\psi(E)dE
\end{equation}
is the definition of the reaction probability integral $\lambda$, that is, the probability per unit time that two nuclei, confined to a unit volume, will react with each other. The reaction probability is written in the significant form $\langle \sigma v\rangle$ to indicate that it is an appropriate average of the product of the reaction cross section and relative velocity of the interacting nuclei. If the reacting mixture is in thermal equilibrium this quantity depends only on the temperature.

\vspace{4mm}

For nonresonant nuclear reactions between nuclei of charges $z_i$
and $z_j$ at low energies (below the Coulomb barrier), the reaction
cross-section has the form (Clayton, 1983, Rowley and Merchant, 1991, Lang, 1999)
\begin{equation}
\sigma(E) = \frac{S(E)}{E} e^{-2\pi \eta(E)}
\end{equation}
where
\begin{equation}
\eta(E) = \left(\frac{\mu}{2}\right)^{1/2} \frac{z_i z_j e^2}{\hbar E^{1/2}}
\end{equation}
is the Sommerfeld parameter, $\hbar$ is Planck's quantum of action,
and $e$ is the quantum of electric charge. Eq. (1.3) allows the extrapolation of measured reaction cross sections down to astrophysical energies by introducing the S-factor.

\vspace{4mm}

It is to be noted that $S(E)$, a residual function of energy,
represents intrinsically nuclear parts of the probability
for the occurrence of a nuclear reaction (Clayton, 1983).  It is
often found to be constant or a slowly varying function over
a limited energy range when the interaction energy of the pair
of nuclei is not nearly equal to an energy at which the
two nuclei resonate in a quasi-stationary state.

\vspace{4mm}

The normalized energy distribution, representing the
isotropic velocity distribution of the reacting nuclei, is given by (Clayton, 1983, Lang, 1999)
\begin{equation}
\psi(E)dE = \frac{2}{\sqrt{\pi}} \frac{E}{kT}
\exp\left(- \frac{E}{kT}\right) \frac{dE}{(kTE)^{1/2}}
\end{equation}
(for reaction rates with anisotropic distributions see Imshennik, 1990).
The substitution of $\psi(E)dE$ in (1.2) yields
\begin{equation}
\lambda = \left(\frac{8}{\mu \pi}\right)^{1/2}
\frac{1}{(kT)^{3/2}}
\int^\infty_0 S(E) \exp\left(-
\frac{E}{kT} - \frac{b}{E^{1/2}}\right)dE.
\end{equation}
where $b=\left(\frac{\mu}{2}\right)^{1/2}\frac{z_iz_je^2}{\hbar}.$
If the residual function $S(E)$ is considered
to be constant, $S_0$, the corresponding reaction probability integral is given by
\begin{equation}
\lambda = \left(\frac{8}{\mu \pi}\right)^{1/2} \frac{S_0}{(kT)^{3/2}}
\int^\infty_0 \exp
\left(- \frac{E}{kT} - \frac{b}{E^{1/2}}\right)dE.
\end{equation}
In effect, the reaction probability is governed by the average of the Gamow penetration factor over the Maxwell-Boltzmann distribution. Anderson et al. (1994) considered for $S(E)$ the Maclaurin
series expansion
\begin{equation}
S(E) = S(0) + \frac{dS(0)}{dE} E +
\frac{1}{2} \frac{d^2 S(0)}{dE^2}E^2
\end{equation}
and solved (1.6) in terms of $G$- and $H$-functions (Mathai and Saxena,
1973, 1978).

\vspace{4mm}

It is to be noted that the residual function $S(E)$ may not
admit the representation (1.8) if the thermonuclear
fusion plasma is not in the thermodynamic equilibrium, that is,
the case in which there is a depletion or cut-off of the high energy tail
of the Maxwell-Boltzman distribution in (1.5) (Lapenta and Quarati, 1993).
Some of these cases have been considered in Anderson et al. (1994) and
Haubold and Mathai (1998).
In this paper we consider representations for the residual function
given by
\begin{eqnarray}
S_1(E) & = & S_0 \delta(E - E_0),\\
S_2(E) & = & S_0 \left(\frac{E}{E_0}\right)^{\alpha-1},\\
S_3(E) & = & S_0\left(\frac{E}{E_0}\right)^{\alpha-1} H(E-E_1),\\
S_4(E) & = & S_0\left(\frac{E}{E_0}\right)^{\alpha-1} H(E_1-E),\\
S_5(E) & = & S_0\left(\frac{E}{E_0}\right)^{\alpha-1}\exp(C(E-E_1)),
\end{eqnarray}
where
\begin{equation}
H(E - E_1):= \left\{\begin{array}{ll}
1, & \hspace*{7mm} \mbox{ if } \; E > E_1,\\[2mm]
0,  & \hspace*{7mm} \mbox{ if } \; E < E_1 \end{array} \right.
\end{equation}
is the unit step function and
\begin{equation}
\delta(E - E_1) := \frac{d}{dE} \left(H(E-E_1)\right)
\end{equation}
is the Dirac delta function (Jennings and Karataglidid, 1998).

\vspace{4mm}

We discuss the analytic representations of the corresponding
reaction probability integrals.  The integral corresponding
to (1.9) may be called instantaneous reaction probability
integral.

\begin{center}
{\bf 2. \, The Astrophysical Thermonuclear Functions}
\end{center}
\renewcommand{\thesection}{2}
\renewcommand{\theequation}{\thesection.\arabic{equation}}
\setcounter{equation}{0}
\vspace{6mm}

The theoretical and experimental verification of nuclear
cross-sections leads to the derivation of the closed-form
representation of the thermonuclear reaction rates. These
rates are expressed in terms of the four
astrophysical thermonuclear functions (given in the notation chosen by
Anderson et al., 1994)
\begin{eqnarray}
&& I_1(z,\nu):   =  \int^\infty_0 y^\nu \exp(-y-\frac{z}{\sqrt{y}})dy,
\\
&& I_2(z,d,\nu):   =  \int^d_0 y^\nu \exp(-y-\frac{z}{\sqrt{y}})dy, \\
&& I_3(z,t,\nu):   =  \int^\infty_0 y^\nu \exp(-y-\frac{z}{\sqrt{y+t}})dy, \\
&& I_4(z,\delta, b,\nu): =  \int^\infty_0 y^\nu \exp(-y-by^\delta
-\frac{z}{\sqrt{y}})dy,
\end{eqnarray}
where $y=\frac{E}{kT}=\frac{\mu v^2}{2kT}$ relates $E$ or $v$, respectively,
to the mean thermal
velocity and $z=2\pi \left(\frac{\mu}{2kT}\right)^{1/2} \frac{z_iz_j e^2}{\hbar}=
2\pi \left(\frac{\mu c^2}{2kT} \right)^{1/2} \alpha z_iz_j$, were the velocity of
light was introduced to make the dimension more apparent and to show the
dependence on Sommerfeld's fine structure constant $\alpha$.
The closed-form representations of these integral functions, in terms of G- and
H-functions, asymptotic values and, numerical results are discussed in
Anderson et al. (1994) and Chaudhry (1999). Accounts about the developments of
Meijer's G-function and of Fox's H-function are given in Mathai and Saxena (1973)
and Mathai and Saxena (1978), respectively. Originally the investigations on these generalized hypergeometric functions were confined to theoretical results such as
their analytical properties, integral representations, and asymptotic expansions,
and then to the study of symmetric Fourier kernels, the solution of certain
functional equations and other mathematical topics. Later, both the G-function and
the H-function have been also used in the fields of statistical and astrophysical
sciences, including extensive studies of the thermonuclear functions in eqs. 
(2.1)-(2.4) (Mathai, 1993).
\vspace{8mm}
\begin{center}
{\bf 3. \, The Extended Gamma Functions}
\end{center}
\renewcommand{\thesection}{3}
\renewcommand{\theequation}{\thesection.\arabic{equation}}
\setcounter{equation}{0}
\vspace{6mm}

The study of the astrophysical thermonuclear functions led to the
development of a new class of special functions (Chaudhry and
Zubair, 1998).  In particular, we define the extended gamma functions by
\begin{equation}
\Gamma(\alpha,x;b;\beta):= \int^\infty_x t^{\alpha-1} e^{-t-b/t^\beta}dt,
\end{equation}
and
\begin{equation}
\gamma(\alpha,x;b;\beta):= \int^x_0 t^{\alpha-1} e^{-t-b/t^\beta}dt.
\end{equation}
It is to be noted that the functions (3.1) and (3.2) are special
cases of the general class of extended gamma functions introduced in
Chaudhry and Zubair (1998).

\vspace{4mm}

In fact we have
\begin{eqnarray}
&& \Gamma(\alpha,x;b;\beta) = \Gamma^{2,0}_{0,2}\left[(b,x)\left|
\begin{array}{cc}
 -, & -  \\
((0,1), & (\alpha,\beta))\end{array} \right.\right] \nonumber\\[2mm]
&& \hspace*{1cm} := \frac{1}{2\pi}
\int^{c+i\infty}_{c-i\infty}\Gamma(s)\Gamma(\alpha+\beta s,x)b^{-s}ds,
\end{eqnarray}
 and
\begin{eqnarray}
&& \gamma(\alpha,x;b;\beta) = \gamma^{2,0}_{0,2}\left[(b,x)\left|
\begin{array}{cc}
 -, & -  \\
((0,1), & (\alpha,\beta))\end{array} \right.\right] \nonumber\\[2mm]
&& \hspace*{1cm} := \frac{1}{2\pi}
\int^{c+i\infty}_{c-i\infty}\gamma(s)\Gamma(\alpha+\beta s,x)b^{-s}ds.
\end{eqnarray}
These functions satisfy the decomposition formula (Chaudhry and
Zubair, 1998).
\begin{equation}
   \gamma(\alpha,x;b,\beta) + \Gamma(\alpha,x;b;\beta) =
  H^{2,0}_{0,2}\left[(b,x) \left|
\begin{array}{cc}
 -, & -  \\
 (0,1), & (\alpha,\beta))\end{array} \right.\right]. \end{equation}
The astrophysical thermonuclear functions (2.1) -- (2.4) are special
cases of the extended gamma functions (3.1) and (3.2) (Chaudhry, 1999)
\begin{eqnarray}
&& I_1(z,\nu) = \Gamma\left(\nu + 1,0;z; \frac{1}{2}\right) \\
&& I_2(z,d,\nu) = \gamma\left(\nu + 1,d;z; \frac{1}{2}\right) \\
&& I_3(z,t,\nu) = e^t \sum^\nu_{r=0} \left(\begin{array}{c} \nu \\ r \end{array} \right)
(-t)^{\nu-t}\Gamma\left(\nu+1, t;z;\frac{1}{2}\right) \\
&& I_4(z,\delta,b,\nu) =   \sum^\nu_{r=0} \frac{(-b)^r}{r!} \Gamma \left(
\nu + r\delta + 1,0;z;\frac{1}{2}\right).
\end{eqnarray}
It is straightforward to note that the transformation theorem
\begin{equation}
\Gamma(\alpha,x;b;\beta) = \frac{1}{\beta} \Gamma^{2,0}_{0,2}
\left[(b^{1/\beta},x)\left|\begin{array}{cc}
- & - \\
\left(\left(0, \frac{1}{\beta}\right), \right. & (\alpha,1)) \end{array} \right.\right],
\hspace*{7mm} (x \geq 0, b \geq 0, \beta \geq 0),
\end{equation}
for the extended gamma function reveals as a special case that
\begin{eqnarray}
&& \Gamma\left(\alpha,0;b;\frac{1}{n}\right) = H^{2,0}_{0,2}\left[b\left|
\begin{array}{cc} -, & - \\
(0,1), & (\alpha,n)\end{array} \right.\right] = (2\pi)^{(1-n)/2}\sqrt{n} \nonumber\\[2mm]
&& \hspace*{1cm} \times \; G^{n+1,0}_{0,n+1}\left[\left(\frac{b}{n}\right)^n\left|\begin{array}{ccc}
-, & -, \ldots, & -\\
0, & \displaystyle \frac{1}{n}, \frac{2}{n},\ldots, \frac{n-1}{n}, & \alpha
\end{array} \right.\right].
\end{eqnarray}
The relation (3.11) yields directly the closed form representation
(Anderson et al., 1994)
\begin{eqnarray}
I_1(z,\nu) & = & \Gamma\left(\nu + 1, 0; z; \frac{1}{2}\right)\nonumber\\[2mm]
& = & \pi^{-1/2}G^{3,0}_{0,3} \left[\frac{z^2}{4}\left|
  0, \frac{1}{2}, 1+\nu   \right.\right].
\end{eqnarray}
The asymptotic representation of the extended gamma function
$\Gamma(\alpha,0;b;\beta)$ for small and large values of $b$ are given
as follows (Chaudhry and Zubair, 1998)
\begin{equation}
\Gamma(\alpha,0;b;\beta) \sim
\left\{\begin{array}{ll}
\displaystyle \Gamma(\alpha) + \frac{1}{\beta} \Gamma\left(- \frac{\alpha}{\beta}\right)
b^{\alpha/\beta}+o(b), & \hspace*{7mm} \mbox{ for small } \;
b, \; \alpha \neq 0, \\
-\displaystyle \frac{1}{\beta}\ln b, & \hspace*{7mm} \mbox{ for small } \; b, \;
\alpha = 0,
\end{array} \right.
\end{equation}
and
\begin{eqnarray}
&& \Gamma(\alpha,0;b;\beta) \sim \frac{1}{\beta}
\left(\frac{2\pi \beta}{1+\beta}\right)^{1/2} \beta^{(2\alpha+\beta)/2(1+\beta)}
\; b^{(2\alpha-1)/2(\beta+1)}\nonumber\\[2mm]
&& \hspace{1cm} \times \; \exp\left[-(1+\beta)^{\beta/(1+\beta)}\;
b^{1/(1+\beta)}\right], \hspace*{7mm} \mbox{ for large } \; b.
\end{eqnarray}
The representations (3.13) and (3.14) are useful in finding the
asymptotic values of the thermonuclear reaction probability integrals.

\vspace{6mm}
\begin{center}
{\bf 4. \, Thermonuclear Reaction Probability Integrals}
\end{center}
\renewcommand{\thesection}{4}
\renewcommand{\theequation}{\thesection.\arabic{equation}}
\setcounter{equation}{0}

\vspace{6mm}

In this section we study the closed form representations of the
thermonuclear reaction probability integrals corresponding to the
various forms (1.9) -- (1.13) of the residual function $S(E)$.

\vspace{4mm}

\noindent {\bf (4.1)\, Instantaneous Residual Function, $S(E) =
S_0 \delta(E-E_0)$}. The substitution of this value of $S(E)$
in (1.6) yields an elementary representation
\begin{equation}
\lambda_1 = \left(\frac{8}{\mu\pi}\right)^{1/2} \frac{S_0}{(kT)^{3/2}}
\exp\left(- \frac{E_0}{kT} - \frac{b}{\sqrt{E_0}}\right)
\end{equation}
of the corresponding thermonuclear probability integral.

\vspace{4mm}

\noindent {\bf (4.b)\, Power Type Residual Function, $
S(E) = S_0(E/E_0)^{\alpha-1}$}. The substitution of the
power type residual function in (1.6) yields
\begin{equation}
\lambda_2 = \left(\frac{8}{\mu\pi}\right)^{1/2}
\frac{S_0}{(kT)^{3/2}}\int^\infty_0 (E/E_0)^{\alpha-1}
\exp\left(- \frac{E}{kT} - \frac{b}{\sqrt{E}}\right)dE.
\end{equation}
The transformation $ t= E/kT$ in (4.2) yields
\begin{equation}
\lambda_2 = \left(\frac{8}{\mu \pi}\right)^{1/2}
\frac{S_0}{\sqrt{E_0}} \left(\frac{kT}{E_0}\right)^{\alpha - \frac{3}{2}}
\int^\infty_0 t^{\alpha-1}\exp\left(- t -
\frac{b}{\sqrt{kT}\sqrt{t}}\right)dt,
\end{equation}
which is available from Haubold and Mathai (1998) and can be written
in terms of the extended gamma function, i.e..
\begin{equation}
\lambda_2 = \left(\frac{8}{\mu \pi}\right)^{1/2}
\frac{S_0}{\sqrt{E_0}} \left(\frac{kT}{E_0}\right)^{\alpha - \frac{3}{2}}
\Gamma\left(\alpha,0;\frac{b}{\sqrt{kT}};\frac{1}{2}\right).
\end{equation}
In view of the decomposition relation (3.5) we can simplify (4.4) to obtain
\begin{equation}
\lambda_2 = \left(\frac{8}{\mu \pi}\right)^{1/2}
\frac{S_0}{\sqrt{E_0}} \left(\frac{kT}{E_0}\right)^{\alpha - \frac{3}{2}}
H^{2,0}_{0,2}\left[\frac{b}{\sqrt{kT}} \left|\begin{array}{cc}
-, & -\\
(0,1), & (\alpha, 2) \end{array} \right.\right]
\end{equation}
that can further be simplified to give
\begin{equation}
\lambda_2 = \left(\frac{8}{\mu\pi}\right)^{1/2}
\frac{S_0}{\sqrt{E_0}}\left(\frac{kT}{E_0}\right)^{\alpha-\frac{3}{2}}
G^{3,0}_{0,3} \left[\frac{b^2}{kT}\left|0, \frac{1}{2},\alpha\right.\right].
\end{equation}

\vspace{4mm}

\noindent {\bf (4.c)\, Power Type Delayed Residual Function,
$S(E) = S_0 (E/E_0)^{\alpha-1}H(E-E_1)$}. The substitution of the
power
type delayed residual function in (1.6) gives
\begin{equation}
\lambda_3 = \left(\frac{8}{\mu\pi}\right)^{1/2}
\frac{S_0}{(kT)^{3/2}} \int^\infty_{E_1}
(E/E_0)^{\alpha-1}\exp\left(- \frac{E}{kT} -
\frac{b}{\sqrt{E}}\right)dE.
\end{equation}
The transformation $t = E/kT$ in (4.7) yields
\begin{equation}
\lambda_3 = \left(\frac{8}{\mu\pi}\right)^{1/2}
\frac{S_0}{\sqrt{E_0}} \left(\frac{kT}{E_0}\right)^{\alpha-\frac{3}{2}}
\int^\infty_{E_1/kT} t^{\alpha-1}\exp\left(-t-
\frac{b}{\sqrt{kT}\sqrt{t}}\right)dt,
\end{equation}
or
\begin{equation}
\lambda_3 = \left(\frac{8}{\mu\pi}\right)^{1/2} \frac{S_0}{\sqrt{E_0}}
\left(\frac{kT}{E_0}\right)^{\alpha-\frac{3}{2}} \Gamma\left(\alpha,
\frac{E_1}{kT};\frac{b}{\sqrt{kT}}; \frac{1}{2}\right).
\end{equation}
The asymptotic representations of $\lambda_3$ for small and
large values of $b/\sqrt{kT}$ can be found from (3.13) and
(3.14).  In fact for small value of $b/\sqrt{kT}$ and $\alpha \neq 0$
\begin{equation}
\lambda_3 \sim \left(\frac{8}{\mu \pi}\right)^{1/2} \frac{S_0}{\sqrt{E_0}}
\left(\frac{kT}{E_0}\right)^{\alpha-\frac{3}{2}}\left\{\Gamma(\alpha)+\frac{2\Gamma(-\alpha)b^2}{kT}
\right\}
\end{equation}

\vspace{4mm}

\noindent {\bf (4.d) \, Power Type Cut-off Residual Function,
$S(E) = S_0(E/E_0)^{\alpha-1}H(E_1-E)$}. The substitution of the
power type cut-off residual function in (1.6) gives
\begin{equation}
\lambda_4 = \left(\frac{8}{\mu\pi}\right)^{1/2} \frac{S_0}{(kT)^{3/2}}
\int^{E_1}_0 (E/E_0)^{\alpha-1}\exp\left(- \frac{E}{kT} -
\frac{b}{\sqrt{E}}\right)dE
\end{equation}
that can similarly be simplified in terms of the extended gamma
function to give
\begin{equation}
\lambda_4 = \left(\frac{8}{\mu\pi}\right)^{1/2}\frac{S_0}{\sqrt{E_0}}
\left(\frac{kT}{E_0}\right)^{\alpha-\frac{3}{2}}\gamma\left(\alpha,\frac{E_1}{kT};
\frac{b}{\sqrt{kT}};\frac{1}{2}\right).
\end{equation}

\vspace{4mm}

\noindent {\bf (4.e)\, Exponential Type Residual Function, \,$
S(E) = S_0(E/E_0)^{\alpha-1} \exp (-C(E-E_1) ),$ $\left(C > -
\frac{1}{kT}\right)$}.
The substitution of the exponential type residual function
in (1.6) yields
\begin{equation}
\lambda_5 = \left(\frac{8}{\mu\pi}\right)^{1/2}\frac{S_0}{(kT)^{3/2}}
\exp(CE_1)\int^\infty_0 (E/E_0)^{\alpha-1}\left(-\left(C + \frac{1}{kT}\right)
E - \frac{b}{\sqrt{E}}\right)dE.
\end{equation}
The transformation
\[
t = \left(C + \frac{1}{kT}\right)E, \hspace*{7mm} \left(C > - \frac{1}{kT}\right)
\]
in (4.13) yields
\begin{equation}
\lambda_5 = \left(\frac{8}{\mu\pi}\right)^{1/2}
\frac{S_0}{(kT)^{3/2}}
\left(\frac{CkT + 1}{kTE_0}\right)^{\alpha-1}\exp(CE_1)
\int^\infty_0 t^{\alpha-1}\exp\left(-t-
\frac{b}{\sqrt{\frac{kT}{CkT+1}}\sqrt{t}}\right)dt.
\end{equation}
The integral in (4.14) is solvable in terms of extended
gamma function to give
\begin{equation}
\lambda_5 = \left(\frac{8}{\mu\pi}\right)^{1/2}\frac{S_0}{(kT)^{3/2}}
\left(\frac{CkT + 1}{kTE_0}\right)^{\alpha-1} \exp(CE_1)
\Gamma\left(\alpha,0;b\sqrt{\frac{kT}{CkT+1}}; \frac{1}{2}\right).
\end{equation}
The equation (4.15) can be simplified further in terms
of the H-function to give
\begin{equation}
\lambda_5 = \left(\frac{8}{\mu\pi}\right)^{1/2}
\frac{S_0}{(kT)^{3/2}}\left(\frac{CkT+1}{kTE_0}\right)^{\alpha-1}
\exp(CE_1) H^{2,0}_{0,2}\left[b{\sqrt{\frac{kT}{CkT+1}}} \left|
\begin{array}{cc}
-, & - \\
\displaystyle (0,1), & (\alpha,2)\end{array} \right.\right].
\end{equation}

\begin{center} 
{\bf 5. \, Conclusion}
\end{center}
\renewcommand{\thesection}{4}
\renewcommand{\theequation}{\thesection.\arabic{equation}}
\setcounter{equation}{0}

\vspace{6mm}

Nuclear reactions govern major aspects of the chemical evolution of
galaxies and stars.  The analytic study of the reaction rates and
reaction probability integrals is important in astrophysics.
The development of the new class of extended gamma functions by
Chaudhry and Zubair has facilitated in convenient notations for
these analytic representations.
We have considered various forms of the residual function occurring
in the reaction probability integrals and have written them analytically
in terms of the extended gamma functions.

\vspace{1cm}

\noindent {\bf Acknowledgment}. The first author is indebted to King Fahd
University of Petroleum and Minerals, Dhahran, Saudi Arabia for excellent
research facilities.

\vspace{1cm}

\begin{center} {\bf References}
 \end{center}

\vspace{6mm}

\baselineskip=14pt
\noindent
Anderson, W.J., Haubold, H.J., and Mathai, A.M.: 1994, Astrophysical
thermonuclear functions, {\em Astrophys. Space Sci.} {\bf 214}, 49--70.
\\[4mm]
Bergstroem, L., Iguri, S., and Rubinstein, H.: 1999, Constraints on the
variation of the fine structure constant from big bang nucleosynthesis, 
{\em Phys. Rev.} {\bf D60}, 045005-1 - 045005-9.
\\[4mm]
Chaudhry, M.A. and Zubair, S.M.: 1998, Extended incomplete gamma
functions with applications, {\em Journal of the London Mathematical
Society} (submitted).
\\[4mm]
Chaudhry, M.A.: 1999, Transformation of the extended
gamma function $\Gamma^{2,0}_{0,2}[(B,X)]$ with applications
to astrophysical thermonuclear functions, {\em Astrophys. Space
Sci.} {\bf 262}, 263--270.
\\[4mm]
Clayton, D.D.: 1983, {\em Principles of Stellar Evolution and
Nucleosynthesis}, Second Edition, The University of Chicago Press,
Chicago and London.
\\[4mm]
Critchfield, C.L: 1972, in: F. Reines (ed.), {\em Cosmology,
Fusion and Other Matters, George Gamow Memorial Volume},
Colorado, Colorado: Associated University Press 1972.
\\[4mm]
Fowler, W.A.: 1984, Experimental and theoretical nuclear astrophysics:
the quest for the origin of the elements, {\em Rev. Mod. Phys.} {\bf 56},
149-179.
\\[4mm]
Haubold, H.J. and Mathai, A.M.: 1998, On thermonuclear reaction rates,
{\em Astrophys. Space Sci.} {\bf 258}, 185--199.
\\[4mm]
Imshennik, V.S.: 1990, Rate of nuclear reactions for an anisotropic
distribution of interacting particles, {\em Sov. J. Plasma Phys.} {\bf 16},
379-383.
\\[4mm] 
Jennings, B.K. and Karataglidid, S.: 1998, $S_{eff}(E)$ and the $^7Be(p, \gamma)^8B$
reaction,\\ http://xxx.lanl.gov/abs/nucl-th/9807007.
\\[4mm]
Lang, K.R.: 1999, {\em Astrophysical Formulae, Vol.I: Radiation, Gas
Processes and High Energy Astrophysics}, Third Enlarged and Revised
Edition, Springer-Verlag, Berlin Heidelberg New York.
\\[4mm]
Lapenta, G. and Quarati, P.: 1993, Analysis of non-Maxwellian fusion reaction
rates with electron screening, {\em Z. Phys}. {\bf A346}, 243-250.
\\[4mm]
Mathai, A.M.: 1993, {\em A Handbook of Generalized Spacial Functions for
Statistical and Physical Sciences}, Clarendon Press, Oxford.
\\[4mm]
Mathai, A.M. and Saxena, R.K.: 1973, {\em Generalized Hypergeometric
Functions with Applications in Statistics and Physical Sciences},
Springer-Verlag, Lecture Notes in Mathematics Vol. 348, Berlin Heidelberg
New York.
\\[4mm]
Mathai, A.M. and Saxena, R.K.: 1978, {\em The H-function with
Applications in Statistics and Other Disciplines}, John Wiley and Sons,
New Delhi.
\\[4mm]
Morel, P., Pichon, B., Provost, J., and Berthomieu, G.: Solar models and
NACRE thermonuclear reaction rates, {\em Astron. Astrophys.} {\bf 350},
275-285.
\\[4mm] 
Rowley, N. and Merchant, A.C.: 1991, Barrier penetration at astrophysical
energies, {\em Astrophys. J.} {\bf 381}, 591-596; \\
(http://pntpm.ulb.ac.be/Nacre/nacre.htm).
\\[4mm]  

\end{document}